\documentclass[journal,twocolumn]{IEEEtran}

\usepackage{cite}
\usepackage[utf8]{inputenc}

\ifCLASSINFOpdf
 \usepackage[pdftex]{graphicx}
\else
\fi

\usepackage{booktabs} 
\usepackage{caption} 
\usepackage{subcaption}
\usepackage[utf8]{inputenc}
\usepackage[english]{babel}
\usepackage{amsmath}
\usepackage{url}
\begin{document}

\title{Deep Learning for Forest Plantation Mapping in Godavari Districts of Andhra Pradesh, India}

\author{Snehal More$^{1,2}$, Anuj Karpatne$^{2}$, Randolph H. Wynne$^{1}$, Valerie A. Thomas$^{1}$ 

\thanks{$^{1}$Department of Forest Resources and Environmental Conservation, Virginia Tech, Virginia, USA. }
\thanks{$^{2}$Department of Computer Science, Virginia Tech, Virginia, USA}
\thanks{*Correspondence: e-mail:snehalm@vt.edu}}

\maketitle

\begin{abstract} small-area forest plantations play a vital role in the socioeconomic well-being of farmers in Southeast Asia. Most plantations are established on former agricultural land, often on land less suitable for agriculture. Plantations that are converted from natural forest have adverse impacts on biodiversity. Mapping small-area plantations is thus important to understand the dynamics of forest cover in Southeast Asia and to study the social, economic, and ecological effects of this important land cover and land use change. While the small size of forest plantations makes it difficult to detect them using moderate resolution satellite sensors, the problem is exacerbated by the high degree of mixing between plantations, surrounding vegetation, and other land covers, which often show variegated responses in satellite signals across space and time. In this paper, we study the problem of mapping small-area forest plantations in East and West Godavari districts of Andhra Pradesh, India using deep learning methods. Remotely sensed cloud-free data from the Harmonized Landsat Sentinel-2 S10 product were classified using a pixel-level neural network and training data labeled using a field-based survey in concert with expert aerial photo interpretation. We compare the performance of deep learning methods with a baseline random forest classifier in our study region of 21543 sq. km over a period of 3 years and analyze the differences in the results across land cover classes and seasons. 
\end{abstract}


\IEEEpeerreviewmaketitle

\section{Introduction}



Abundant availability of remotely sensed data from satellite products such as MODIS, Landsat, Sentinel, SRTM, NAIP, LISS-3 and AWiFS provide unprecedented opportunities for mapping forest cover using satellite-based images in a timely and cost-effective manner. In recent years, a number of environmental studies have  been conducted that mapped forest cover using remotely sensed data ranging from moderate to high spatial resolutions. For example, a global forest cover change study was performed in \cite{hansen} using Landsat data at 30 m resolution. A number of local and regional scale studies have also been conducted in different parts of the world, such as a study of deforestation and forest fragmentation in India using Landsat and AWiFS data\cite{paper1}, and vegetation cover mapping of India using AWiFS data\cite{paper2}. 

\begin{figure}[!ht]
    \centering
    \includegraphics[width=85mm]{{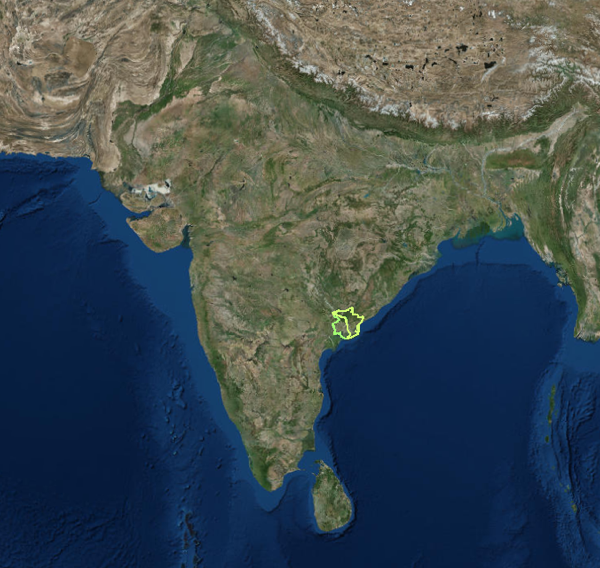}}
    \caption{Study Area: East Godavari and West Godavari  Districts}
    \label{fig:Study Area}
\end{figure}

In our project, we focus on the East and West Godavari districts in Andhra Pradesh, India as our region of study (see Figure \ref{fig:Study Area}). Natural forest cover has been decreasing in the study area in recent years \cite{news_link}. Simultaneously, however, new small-area plantations (average size less than 2 ha) have emerged mainly from conversion of degraded agricultural land to forest plantations. Unfortunately, detection of these plantations using remotely sensed data has been constrained due to multiple factors like the ultra-small size of the farms compared to publicly available moderate resolution satellite images, e.g., from Landsat and MODIS, limited number of accurately-labeled landcover samples, short rotation cycle of plantations, and mixing of plantation with surrounding cropland that has similar spectral reflectance values in both time and space.

This study aims to detect and monitor small-area forest plantations, which, as per local terminology, can be defined as trees outside natural forests, with improved accuracy in our region of study. The broader goal of our project is to analyze and understand the socioeconomic factors that drive decision making of low- to medium- range income farmers in choosing plantation over other crops\cite{NASA_project}. We use the Harmonized Landsat Sentinel-2 (HLS) S10 imagery at 10 m, 20 m and 60 m spatial resolutions\cite{HLS} for our study. 

Automated analysis of high-resolution satellite data using machine learning methods is typical among remote sensing practitioners. However, most operational efforts use random forests, kNN, or classification and regression trees. In this study, we use a pixel-based deep learning (DL) classifier trained using training data obtained from a field-based survey, with a focus on separating natural forests, small-area forest plantations and other croplands.  We compare the classification accuracy of our DL model with the commonly used random forests classifier in our region of study, and analyze the variations in performance across different land covers and seasons.

The remainder of the paper is organized as follows. Section 2 provides a summary of the satellite data used in this study for training and validation. Section 3 provides an overview of the deep learning model used. Section 4 discusses the preliminary results of our study while Section 5 provides concluding remarks and future directions of research.

\section{Data}
For this study, we used satellite images and point location based land-cover labeled data for training and validation. Remotely sensed S10 image data from Harmonized Landsat and Sentinel-2 (HLS) series was used for the land cover analysis. The HLS products are based on a set of algorithms to obtain seamless products from both sensors, i.e., Landsat and Sentinel-2. S10 consist of 10 m, 20 m and 60 m spatial resolutions at various spectral bands. HLS images from November, 2015 to October, 2018 were provided by NASA. Our study area includes East Godavari and West Godavari districts spread across 6 HLS tiles (44QME, 44QMD, 44QNE, 44QND, 44QPE, 44QPD). 

 The image band value extraction for a given spatial location was done using the MATLAB hdf tool. Pursuant to typical practice in multitemporal change detection studies \cite{brooks}, we used every cloud-free observation. The 8bit Quality Assessment (QA) band was used to mask cloud-covered pixels. Each labeled observation consists of the following input variables for a given point location on the ground:  14 spectral reflectance band values (including the QA band) and a temporal variable (Day of Year) to account for seasonal variations in land cover classification performance (discussed in detail in Section 4).
 
\begin{figure}[!ht]
    \centering
    \includegraphics[width=85mm]{{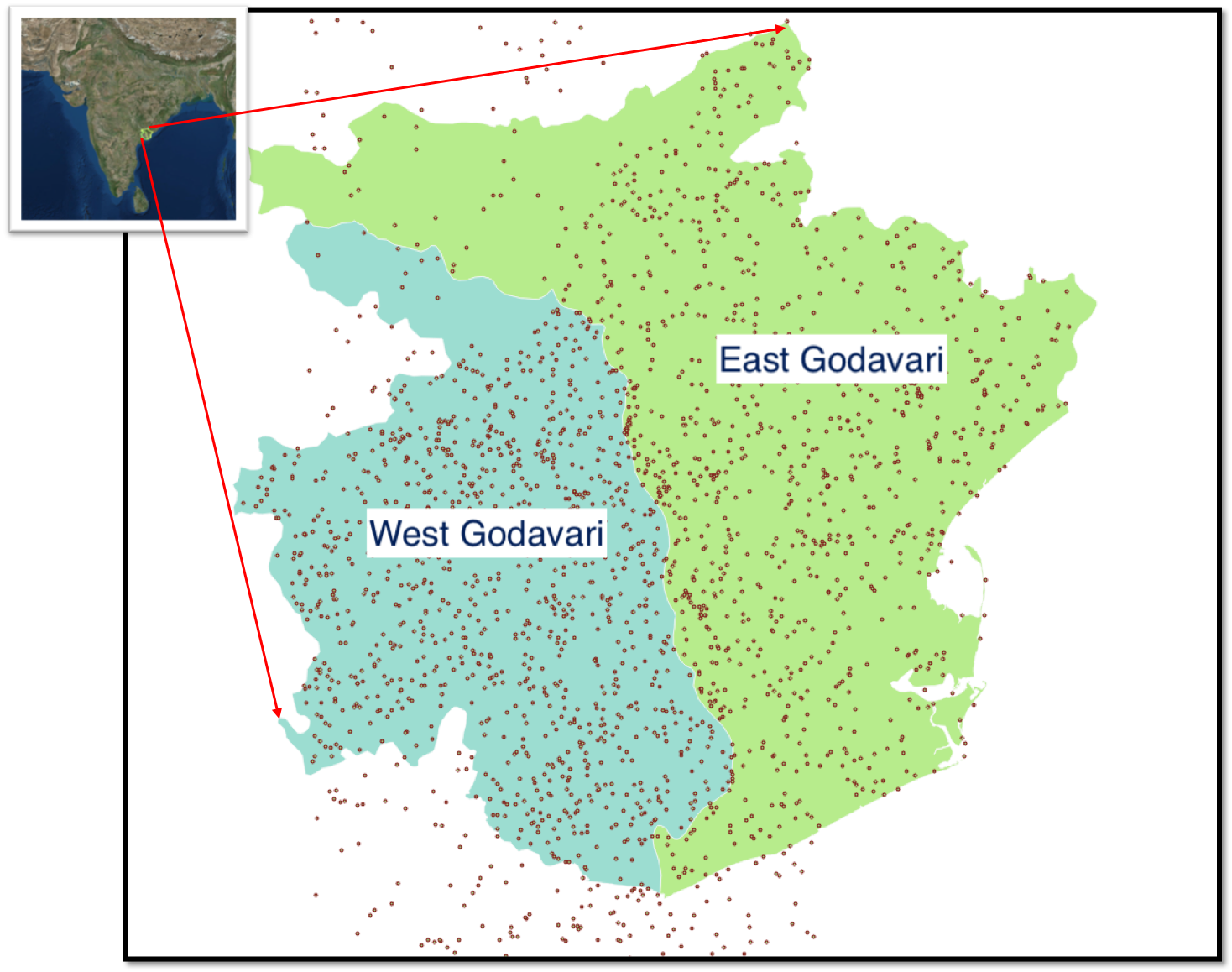}}
    \caption{Spatial distribution of labeled land-cover points over study area}
    \label{fig:Points}
\end{figure}

\begin{figure}[!ht]
    \centering
    \includegraphics[width=85mm]{{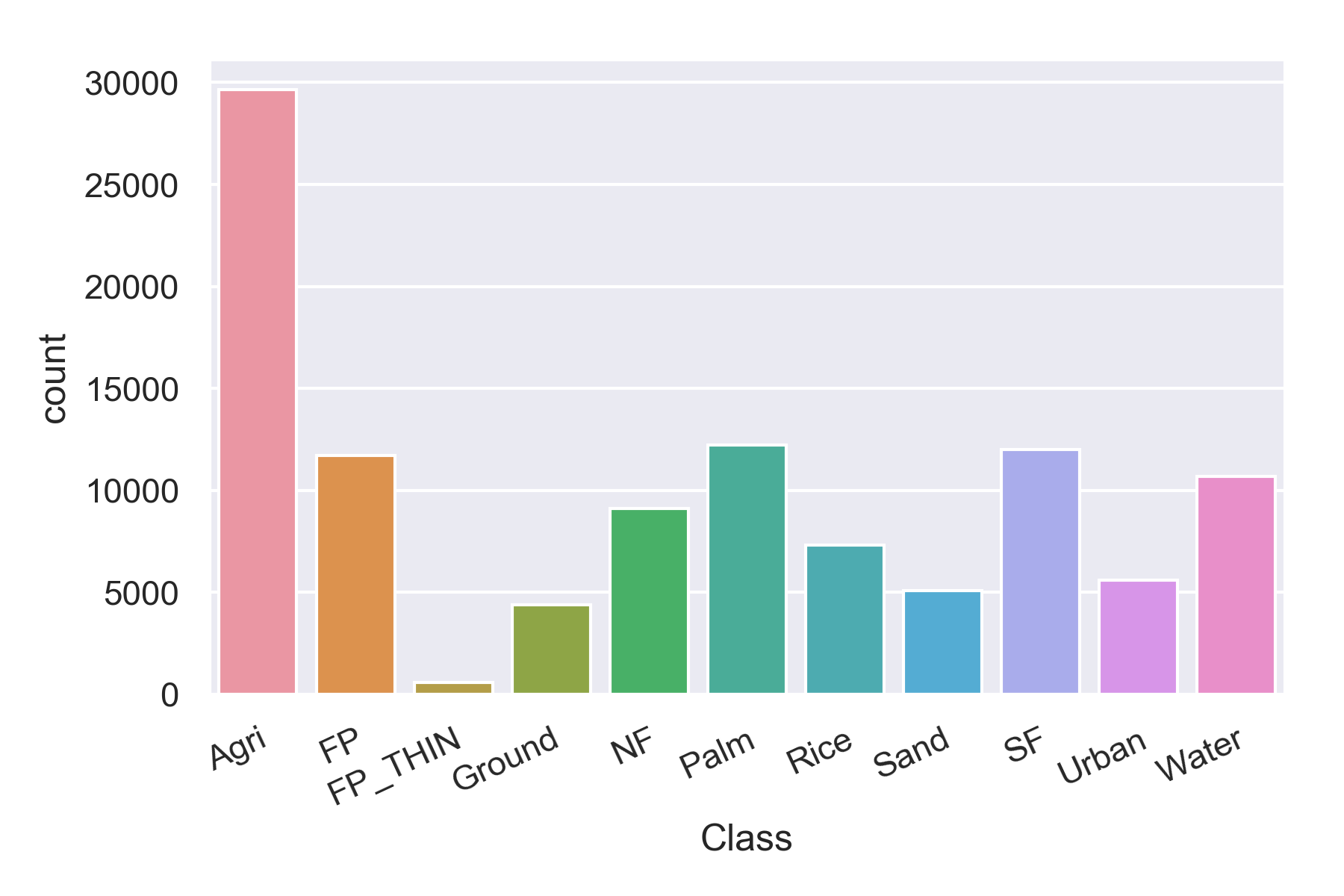}}
    \caption{Class frequency distribution of available labeled dataset}
    \label{fig:frequency}
\end{figure}

For the purpose of training and validation, ground-truth labels at 2276 point locations randomly distributed over our study area were collected by a Virginia Tech student through visual interpretation (Google Earth), categorized into the following 11 land cover classes: 
agriculture (excluding rice), forest plantation (FP), forest plantation thinned (FP$_\text{THIN}$), bare ground, natural forest, palm (oil palm and coconut palm), rice crop, sand, scattered forest (SF), urban and water. Figure \ref{fig:Points}) shows the spatial distribution of labeled points as magenta-colored dots over the study area. These points were chosen such that they in the middle of a single land cover type to avoid mixed pixels. 
https://www.overleaf.com/2272115428zbcqmmymsmgf
 As the points were collected randomly over the study area, the class frequency distribution is not even. Figure \ref{fig:frequency} shows the distribution of class frequencies in our labeled data set. We can see that the classes of interest for forest plantation mapping, i.e., FP, FP$_\text{THIN}$, NF, and Palm, are skewed and less frequent as compared to the Agriculture class.
 
\begin{figure}[!ht]
    \centering
    \includegraphics[width=75mm]{{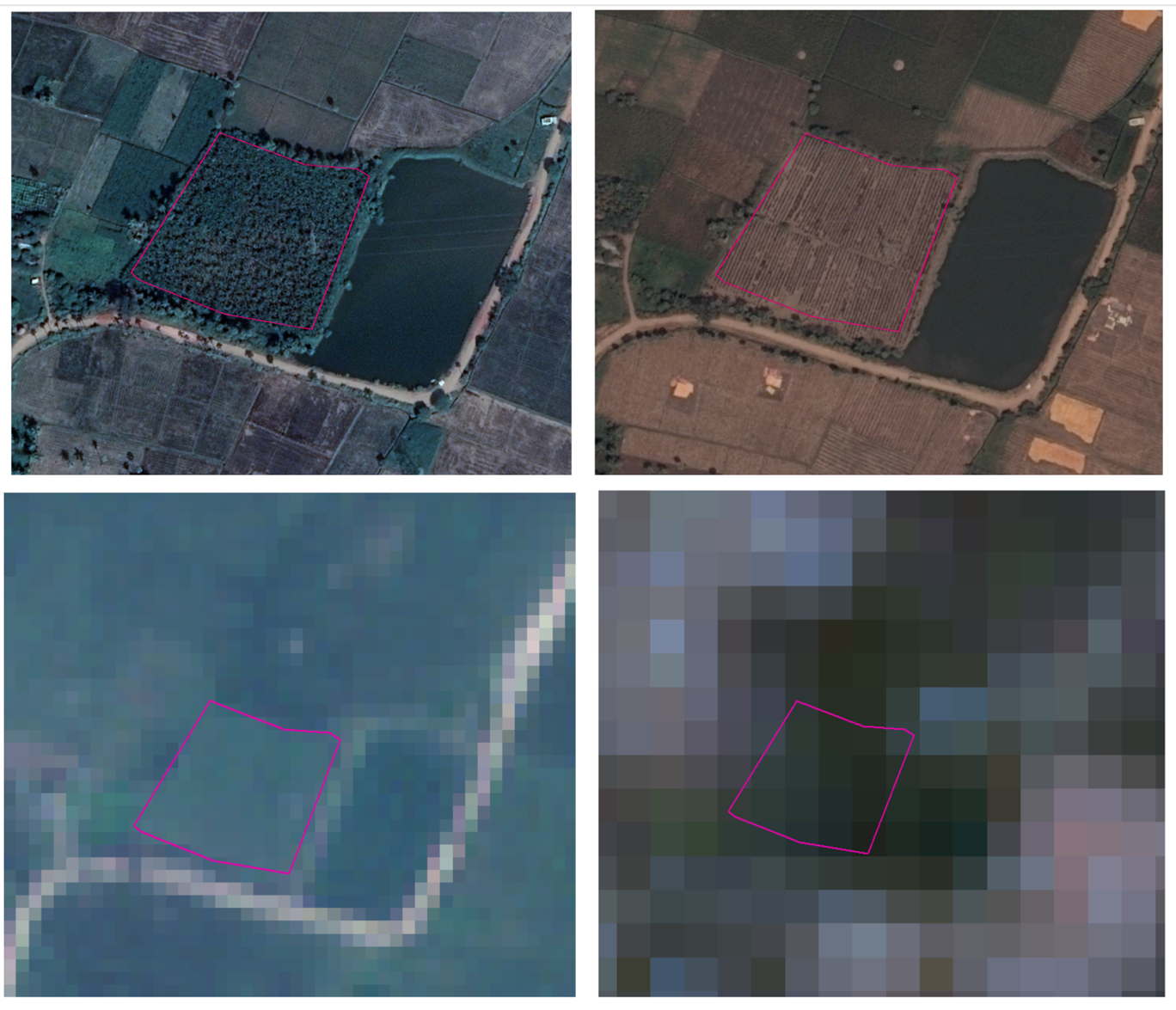}}
    \caption{Sample Casuarina plantation with image backdrops of different spatial resolutions. The first row is the VHR data from Google Earth (December 2017 and March 2016) with Sentinel VNIR at 10m (left) and Landsat at 30m (right) shown on the second row}
    \label{fig:resolutions}
\end{figure}

It is worth mentioning that while HLS provides data at a moderately high spatial resolution (10 m to 60 m), it is still coarse for mapping small-area plantations and agricultural plots in our region of study. We found that the average agricultural plot size in our study area is less than 2 ha (some plot sizes are as small as 0.5 ha). This corresponds to 50 to 200 number of pixels per plot in HLS data at 10 m resolution, which is quite low for building spatial classifiers such as convolutional neural networks that extract region-level features from spatially contiguous groups of pixels. Figure \ref{fig:resolutions} shows an example of a Casuarina plantation observed in Sentinel and Landsat data (bottom row), which is inadequately coarse compared to very-high resolution data from Google Earth (top row). As a result, we adopted a pixel-based deep learning classifier for our current study involving HLS data. With access to higher resolution data in our region of study at 3 m, e.g., from  Planet or Worldview(based on availability), we plan to develop spatial classifiers in a follow-on study.

 
\section{Methodology}

\begin{figure}[h!]
    \centering
    \includegraphics[width=50mm]{{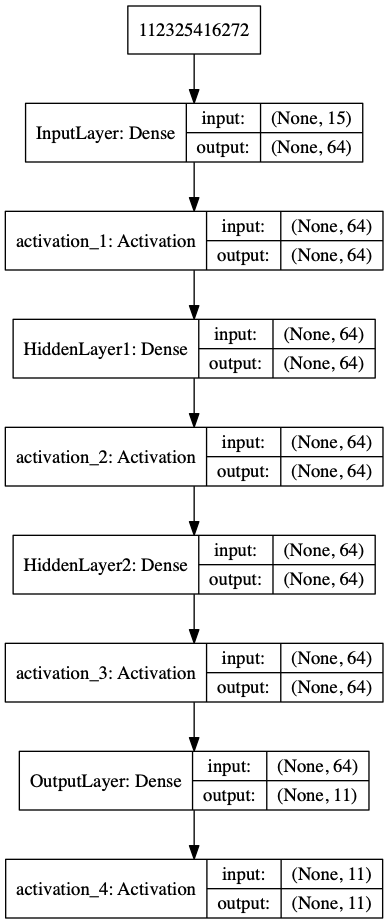}}
    \caption{Fully connected Feed-forward Neural Network Architecture}
    \label{fig:model_plot}
\end{figure}

A fully connected feed-forward deep learning model was used in our study for pixel-level land cover classification using HLS data, available every 3 to 8 days in our 3-year study period from November, 2015 to October, 2018. We built a Keras sequential neural network with ReLU and Softmax as activation functions, Adam as Optimizer, and categorical-crossentropy as loss function (see Figure \ref{fig:model_plot} for the complete architecture). This architecture has 15 input nodes corresponding to 14 band values and DoY (Day of Year) as the 15$^{th}$ input variable. Dense fully connected hidden layers consist of 64 units each, the output layer consists of 11 units corresponding to 11 output classes (Agriculture, Forest Plantation, Forest Plantation Thin, Ground, Natural Forest, Palm, Rice crop, Sand, Scattered Forest, Urban, Water). The Indian subcontinent has three cropping cycles in a year, namely Rabi, Kharif and Summer crops. It was observed during our field visit to the study area that most of the farmers plant and harvest 2 to 3 crops in a year. On the other hand, natural forest and forest plantations have a longer harvest cycle. 

 \begin{figure}[!ht]
    \centering
    \includegraphics[width=70mm]{{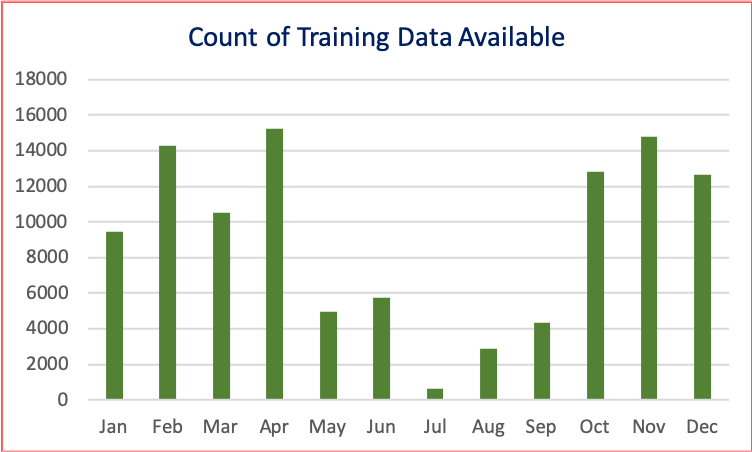}}
    \caption{Cloud-free dates per month summed across all sample locations}
    \label{fig:Monthly_data}
\end{figure}

Each monthly dataset was split into two sets for training and validation using stratified random sampling where we maintain an equal percentage of each class in training and testing datasets. Each month’s model was trained and validated separately. We used the remaining half of the data which was not seen by the model previously for the purpose of testing. 

Figure \ref{fig:Monthly_data} provides a distribution of the number of cloud-free observations of the training data in our region of study for different months of a year. For the monsoon season months from July to October, there were fewer cloud free observations. The July model results are not considered in our analysis as the number of training samples for one of the classes was fewer than ten.


\section{Result and Observations}

\begin{figure}[ht]
    \includegraphics[width=85mm]{{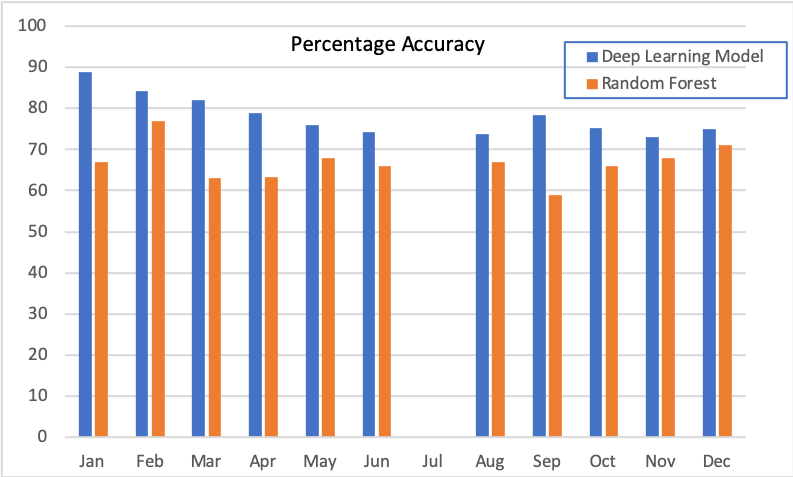}}
\caption{Overall classification accuracy for RF and DL model}
\label{fig:accuracy}
\end{figure}

Figure \ref{fig:accuracy} compares the classification accuracy of our deep learning model for every month in the year. We can observe that the accuracy is lower in August due to the paucity of training data. The accuracy is highest in January and February because of the relatively large number of cloud-free observations and easier separability among the classes. We also compare the performance of our deep learning model with a baseline random forests consisting of 200 decision trees in Figure \ref{fig:accuracy} and observe consistent improvements in accuracy across all seasons of the year. 
\begin{figure}[h!]
    \includegraphics[width=90mm]{{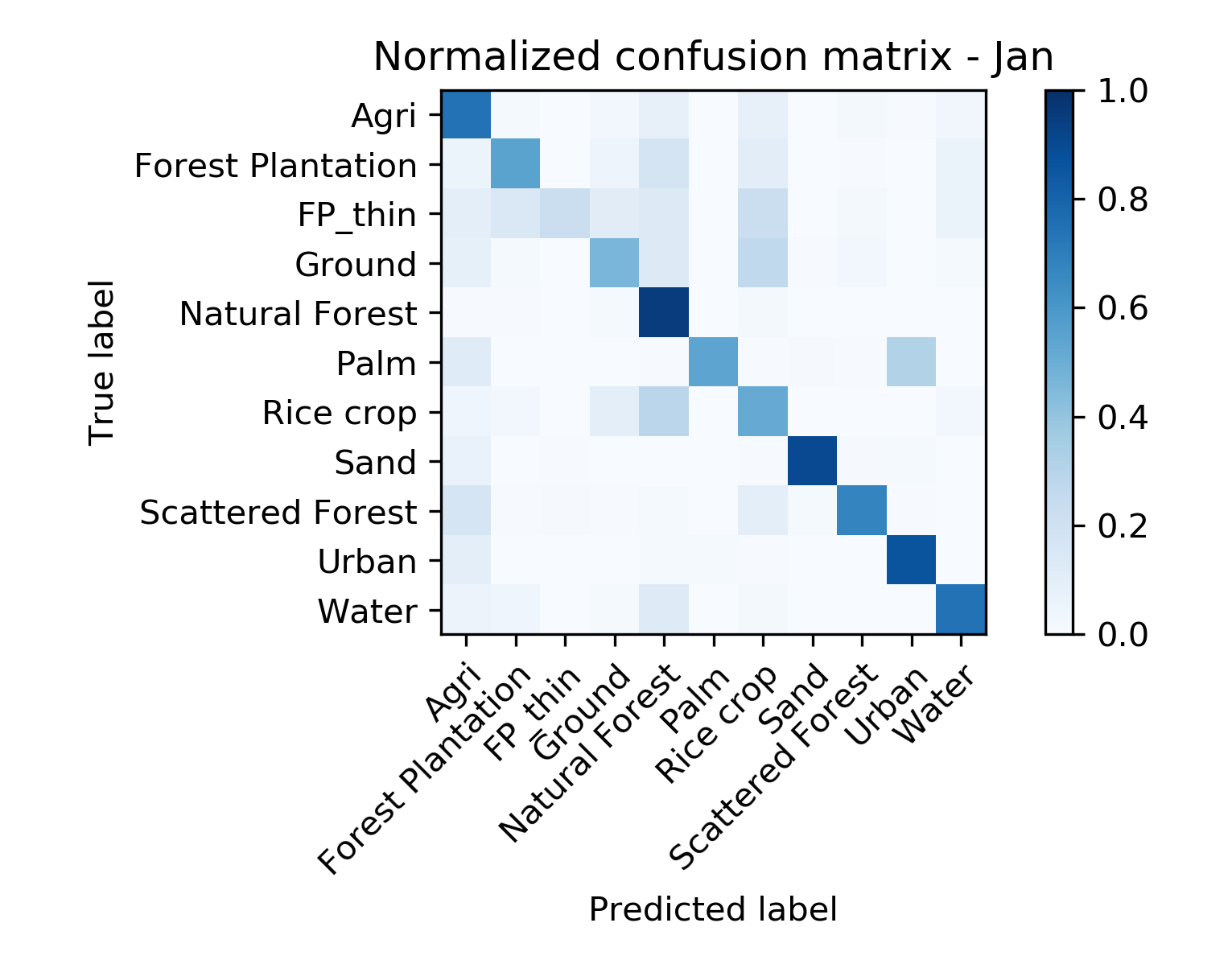}}
    \includegraphics[width=90mm]{{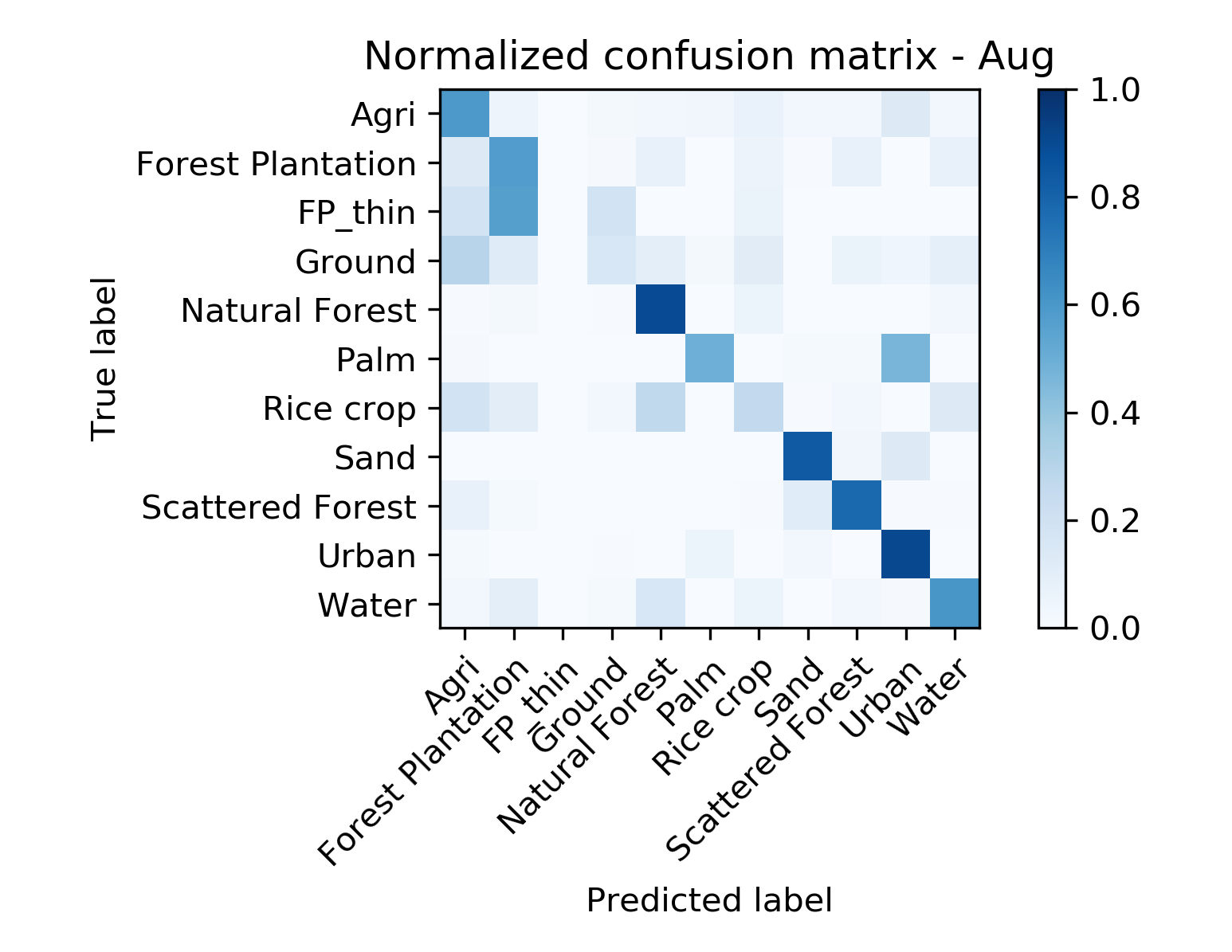}}
\caption{Normalized confusion matrices for our deep learning classification models in the months of January and August}
\label{fig:confusion_mat}
\end{figure}



\begin{figure}[h!]
    \includegraphics[width=80mm]{{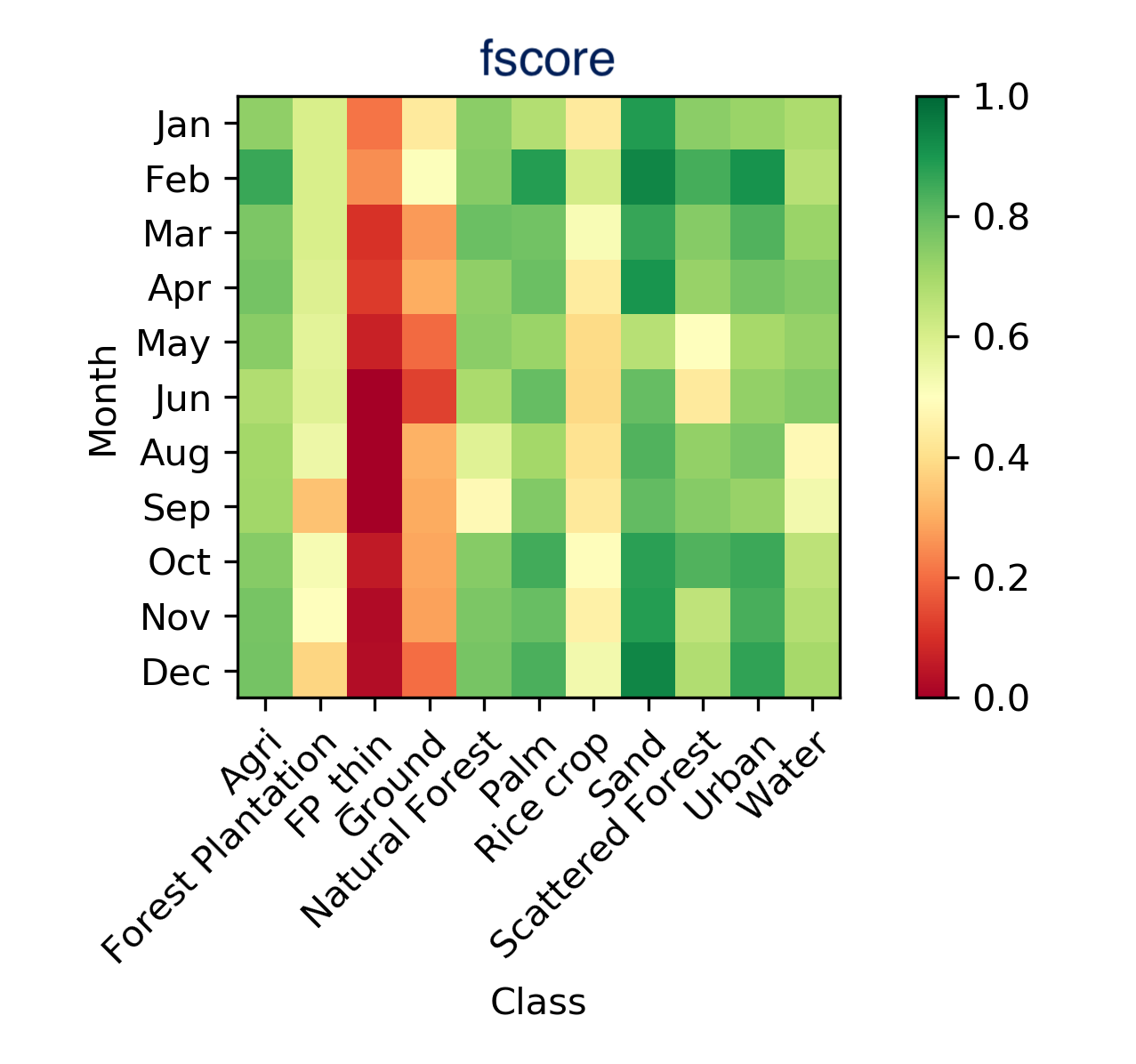}}
\caption{F-score heat map for DL classification model}
\label{fig:NN_fscore}
\end{figure}
\begin{figure}[h!]
    \includegraphics[width=80mm]{{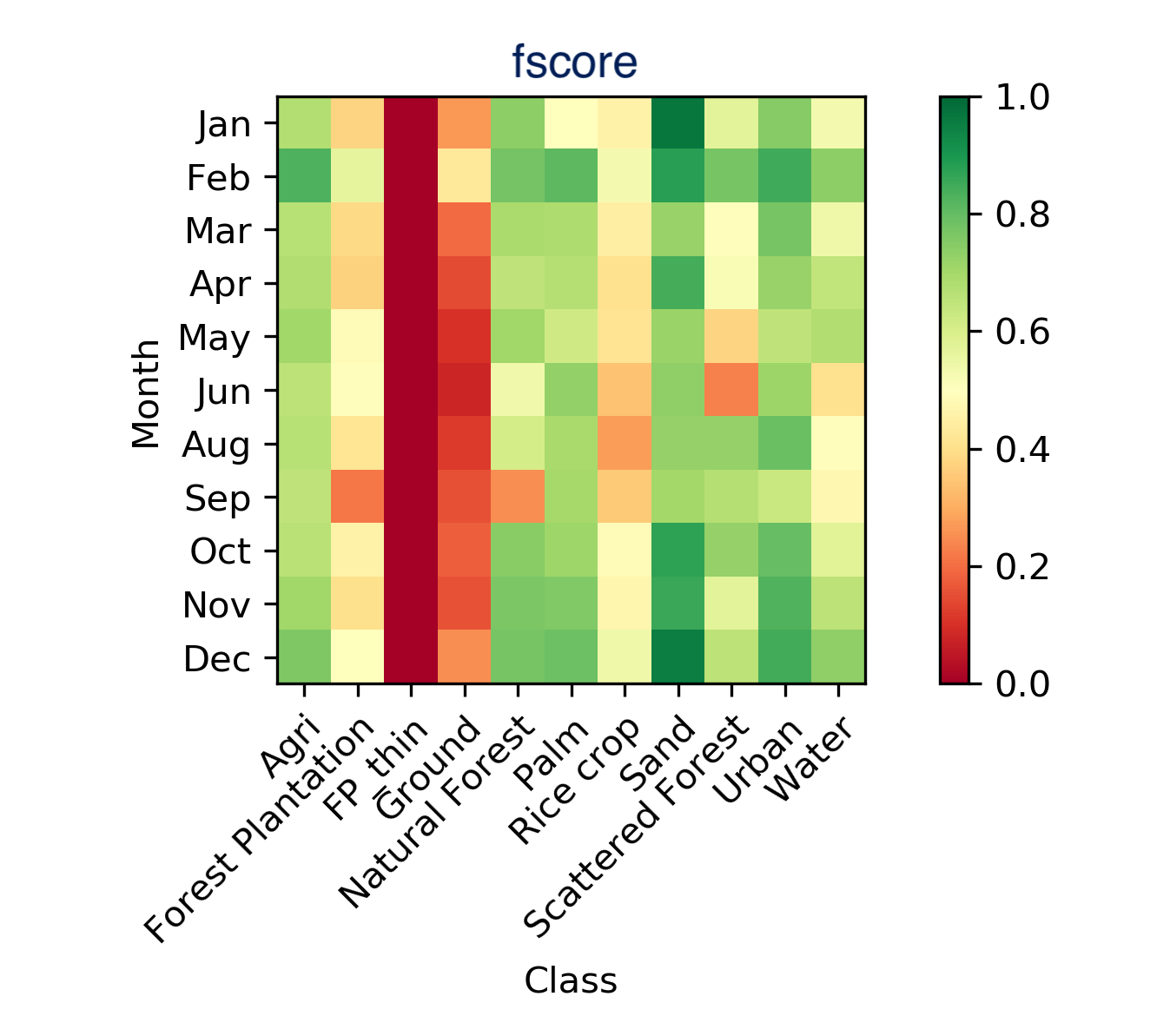}}
\caption{F-score heat map for RF  classification model}
\label{fig:RF_fscore}
\end{figure}

 Since the overall accuracy across all classes may not be a suitable evaluation metric in the presence of skewed class distributions, we provide a comparison of class-wise F-measure scores for our deep learning model and random forests for every month in Figures \ref{fig:NN_fscore} and \ref{fig:RF_fscore}, respectively. We can see that the two most skewed classes (with the smallest relative number of training points in Figure \ref{fig:frequency}), ``FP thin'' and ``Ground,'' have have the maximum F-scores for both classifiers, especially in the monsoon months. However, our deep learning model shows superior F-measures across all seasons for both these classes, in comparison with random forests.

In order to better understand the differences in classification performance and the pair-wise confusion among land cover types, Figure \ref{fig:confusion_mat} provides the confusion matrices normalized w.r.t. to the true class labels (represented as rows in the matrix) for our deep learning approach for two representative months of the year: January and August (Figure \ref{fig:confusion_mat}). We observe in Figure \ref{fig:NN_fscore} and Figure \ref{fig:RF_fscore} that the confusion between bare ground and rice crop varies between January and May. Sand classification accuracy is higher throughout the year, whereas thinned forest plantation has maximum confusion with other classes. Natural forests are well-classified, but the classification accuracy for forest plantation is relatively low due to the issues earlier raised (small size, spectral similarity to agriculture, etc.).

\section{Conclusions and Future Work}

While our current analysis provides a preliminary view of the differences in classification accuracy between deep learning and random forests, future work can focus on an in-depth evaluation of the difference between the methods using comprehensive parameter tuning experiments.  

Moving forward, the number of classes will be revised to correspond to an established land-cover classification scheme. In a subsequent study, we plan to classify Planet imagery using convolutional neural network models followed by multi-source transfer learning analysis. The land cover model from a concomitant econometric analysis  
will also be used to guide deep learning model development. 

\section*{Acknowledgment}
We thank International Paper for providing the locations of known source plantations. We also thank Ms. Paige Williams for the collection and provision of the remaining data points used in this study.



\begin{thebibliography}{1}

\bibitem{NASA_project}
Spatiotemporal Drivers of Fine-Scale Forest Plantation Establishment in Village-Based Economies of Andhra Pradesh\\
\url{http://sari.umd.edu/content/spatiotemporal-drivers-fine-scale-forest-}
\url{plantation-establishment-village-based-economies}\\

\bibitem{HLS}
The Harmonized Landsat and Sentinel-2 surface reflectance data set\\
\url{https://www.sciencedirect.com/science/article/pii/S0034425718304139?via%3Dihub}\\

\bibitem{hansen}
High-Resolution Global Maps of 21st-Century Forest Cover ChangeHigh-Resolution Global Maps of 21st-Century Forest Cover Change \\ by Hansen, M. C., Potapov, P. V., Moore, R., Hancher, M., Turubanova, S. A., Tyukavina, A., Thau, D., Stehman, S. V., Goetz, S. J., Loveland, T. R., Kommareddy, A., Egorov, A., Chini, L., Justice, C. O., Townshend, J. R. G.\\ 
\url{https://science.sciencemag.org/content/342/6160/850/tab-pdf}\\

\bibitem{paper1}
National assessment of forest fragmentation in India: Landscape indices as measures of the effects of fragmentation and forest cover change\\
by Sudhakar Reddy, S. Sreelekshmi, C.S.Jha, V.K.Dadhwal
\url{https://www.sciencedirect.com/science/article/abs/pii/S0925857413004102}\\ 


\bibitem{paper2}
Vegetation cover mapping in India using multi-temporal IRS Wide Field Sensor (WiFS) data\\
by P.K. Kumar  Joshi, P.S.Roy,  Sarnam Singh, Shefali Agrawal, Deepshikha Yadav\\
\url{https://www.sciencedirect.com/science/article/abs/pii/S0034425706001520}\\

\bibitem{news_link}
India's forest cover declines, blame on Naxals | Andhra Pradesh News \\
\url{https://www.sciencedirect.com/science/article/abs/pii/S0034425706001520}\\

\bibitem{brooks}
Fitting the multitemporal curve: A fourier series approach to the missing data problem in remote sensing analysis\\
by Brooks, Evan B., Thomas, Valerie A., Wynne, Randolph H., Coulston, John W.\\
\url{https://ieeexplore.ieee.org/abstract/document/6155734}\\



\end{thebibliography}
\end{document}